\documentclass[aps,pra,twocolumn,showpacs,superscriptaddress,floatfix]{revtex4}
\usepackage{graphicx}
\usepackage{bm}
\usepackage{amsmath}
\usepackage{txfonts}
\usepackage{subfigure}
\begin{document}
\title{Conical intersections in laboratory coordinates with
ultracold molecules}
\author{Alisdair O. G. Wallis}
\affiliation{Department of Chemistry, Durham University,
South Road, Durham, DH1~3LE, United Kingdom}
\author{S. A. Gardiner}
\affiliation{Department of Physics, Durham University,
South Road, Durham DH1~3LE, United Kingdom}
\author{Jeremy M. Hutson}
\affiliation{Department of Chemistry, Durham University,
South Road, Durham, DH1~3LE, United Kingdom}

\date{\today}

\begin{abstract}
For two states of opposite parity that cross as a function of
an external magnetic field, the addition of an electric field
will break the symmetry and induce an avoided crossing. A
suitable arrangement of fields may be used to create a conical
intersection as a function of external spatial coordinates. We
consider the effect of the resulting geometric phase for
ultracold polar molecules. For a Bose-Einstein condensate in
the mean-field approximation, the geometric phase effect
induces stable states of persistent superfluid flow that are
characterized by half-integer quantized angular momentum.
\end{abstract}


\maketitle

%\fbox{\parbox{0.95\linewidth}{{\em Note for copy editor:} We have
%been very careful to make correct use of Roman and italic
%subscripts and superscripts, with Roman for abbreviations and
%italic for mathematical indices. Please do not change all our
%subscripts and superscripts to italic.}}

It is well known that the potential energy surfaces for
molecular electronic states of the same symmetry can cross at a
point in 2 dimensions or on a surface of dimension $n-1$ in $n$
dimensions. These crossings are known as conical intersections
because the two surfaces locally form a double cone. Conical
intersections have a wealth of interesting consequences for
molecular structure and dynamics. For example, they are
responsible for the Jahn-Teller effect
\cite{Longuet-Higgins:1958} and play an important role in
nonadiabatic processes \cite{Domcke2004}. One of the most
interesting consequences of conical intersections is the {\em
geometric phase} (Berry phase) effect \cite{Berry:1984}: when
the nuclei follow a path that encircles a conical intersection
once and returns to the original configuration
(pseudorotation), the electronic wavefunction changes sign.
Since the {\em total} wavefunction must be a single-valued
function of coordinates, this requires that the wavefunction
for nuclear motion must also change sign. This has important
dynamical consequences: it produces half-odd-integer
quantization for free pseudorotation \cite{Busch1998} and may
have significant effects on collision cross sections
\cite{Juanes-Marcos:2005}.

The purpose of the present paper is to explore conical
intersections of a different type. It is now possible to
produce atomic and molecular Bose-Einstein condensates (BECs)
and to subject them to applied magnetic and electric fields.
The atomic and molecular states split and shift as a function
of the magnetic field (Zeeman effect) and electric field (Stark
effect). In the absence of an electric field, parity is
conserved, so it is possible to tune the magnetic field so that
two levels of different parity are exactly degenerate with one
another. However, if a simultaneous electric field is applied,
the two levels of opposite parity are mixed and the degeneracy
is resolved \cite{Friedrich:2000}. Conical intersections can
thus occur at points where the electric field is zero. It is
possible to envisage an arrangement of fields that creates
conical intersections between two atomic or molecular levels
{\em as a function of external spatial coordinates} rather than
internal coordinates.

All the results that apply to conical intersections between
molecular potential energy surfaces will continue to apply in
this new situation. The internal (electronic/vibrational/spin)
wavefunction of the atom or molecule will change sign along a
path that encircles the intersection once, so the spatial
wavefunction of the condensate must also change sign.

Quantized vortices are a characteristic sign of superfluid flow
\cite{annett2004}, and since their first observation in the
context of Bose-Einstein condensed dilute atomic gases
\cite{matthews1999}, have been key to some spectacular
experimental results \cite{raman2001,leanhardt2002}. Recent
experiments placing a BEC within a toroidal trapping geometry
\cite{ryu2007, arnold2006, henderson2009} have enabled the
observation of persistent flow around a toroidal trap
\cite{ryu2007}. The possibility of forming half-integer
quantized vortices within a spinor atomic BEC configuration has
been investigated theoretically \cite{zhou2001,
ruostekoski2003, chiba2008, hoshi2008}, as have a variety of
differing consequences of geometric phase effects in atomic BEC
systems \cite{fuentes-guridi2002, chen2004, zhang2006,
kaurov2006}. In this work we combine these different threads to
show how, with an appropriate configuration of magnetic and
electric fields, a BEC of \textit{heteronuclear diatomic
molecules} will assume a toroidal geometry, such that the
geometric phase causes the system to manifest macroscopically
occupied states of half-integer quantized persistent flow.

Effects of this type can in principle be observed in any system
where two states of opposite parity can be tuned into
degeneracy with a magnetic field and can be coupled with an
electric field. However, for atomic systems states of different
parity are usually far apart at zero field. More accessible
examples are provided by gases of heteronuclear alkali-metal
dimers such as RbCs and KRb, which are the targets of current
experiments. In the present work we illustrate the effect for a
gas of KRb molecules in a single vibrational level of the
lowest triplet state, $^3\Sigma^+$.

The energy levels of a $^3\Sigma^+$ molecule in an applied field
are conveniently expanded in a fully decoupled basis set of
functions $|NM_N\rangle |SM_S\rangle$, where $N$ and $S$ are
quantum numbers for molecular rotation and electron spin and $M_N$
and $M_S$ are the corresponding space-fixed projections onto the
magnetic field axis. Nuclear spin is neglected here for
simplicity. A simple form of the Hamiltonian that contains all the
essential ingredients is
\begin{equation}
H = B\hat N^2 + \frac{2}{3}\lambda\left(\hat S^2 - 3\hat
S_Z^2\right) + g_e \mu_{\rm B} B_Z M_S - \bm{\mu}.\bm{E},
\end{equation}
where $B$ is the molecular rotational constant, $\lambda$ is
the spin-spin coupling constant, $S_Z$ is the projection of $S$
onto the molecular axis, $B_Z$ is the magnetic field orientated
along the space-fixed $Z$ axis, and $\bm{E}$ and $\bm{\mu}$ are
the vectors representing the electric field and molecular
electric dipole moment. KRb has not been characterized in
detail spectroscopically but ultracold KRb has recently been
formed in the lowest rovibrational levels of both singlet and
triplet states \cite{Ni:KRb:2008}. Electronic structure
calculations give an equilibrium distance $r_{\rm
e}=5.901$~\AA\ for the triplet state \cite{Rousseau:2000}. This
allows $B$ and $\lambda$ for the lowest vibrational level to be
estimated as $B=0.01813$ cm$^{-1}$ and $\lambda=-0.00632$
cm$^{-1}$. The dipole moment function has been calculated by
Kotochigova, Julienne and Tiesinga \cite{Kotochigova:2004} and
has a value around 0.051~D near $r_{\rm e}$.
\begin{figure}[tb]
\includegraphics[width=65mm]{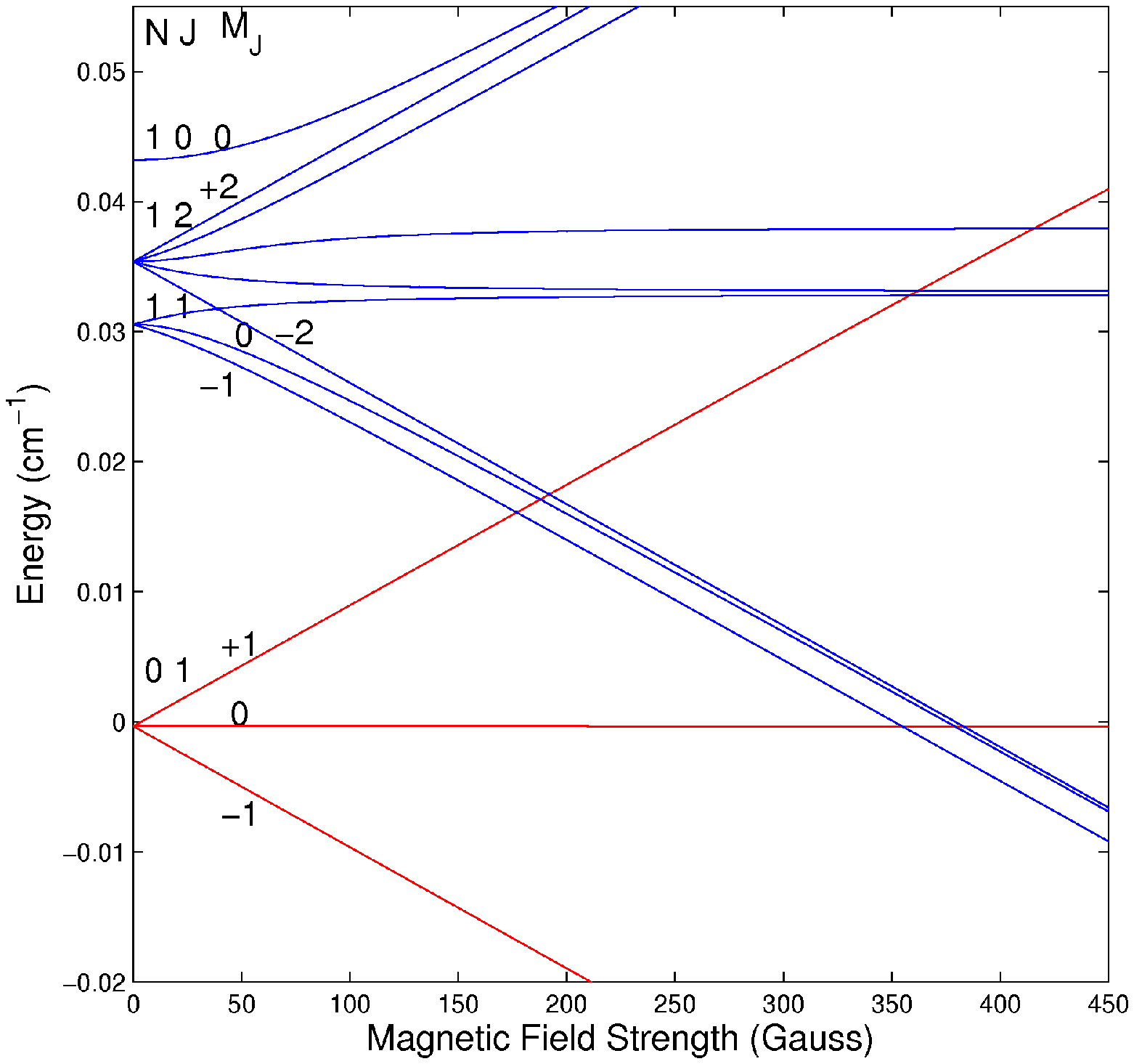}
\includegraphics[width=65mm]{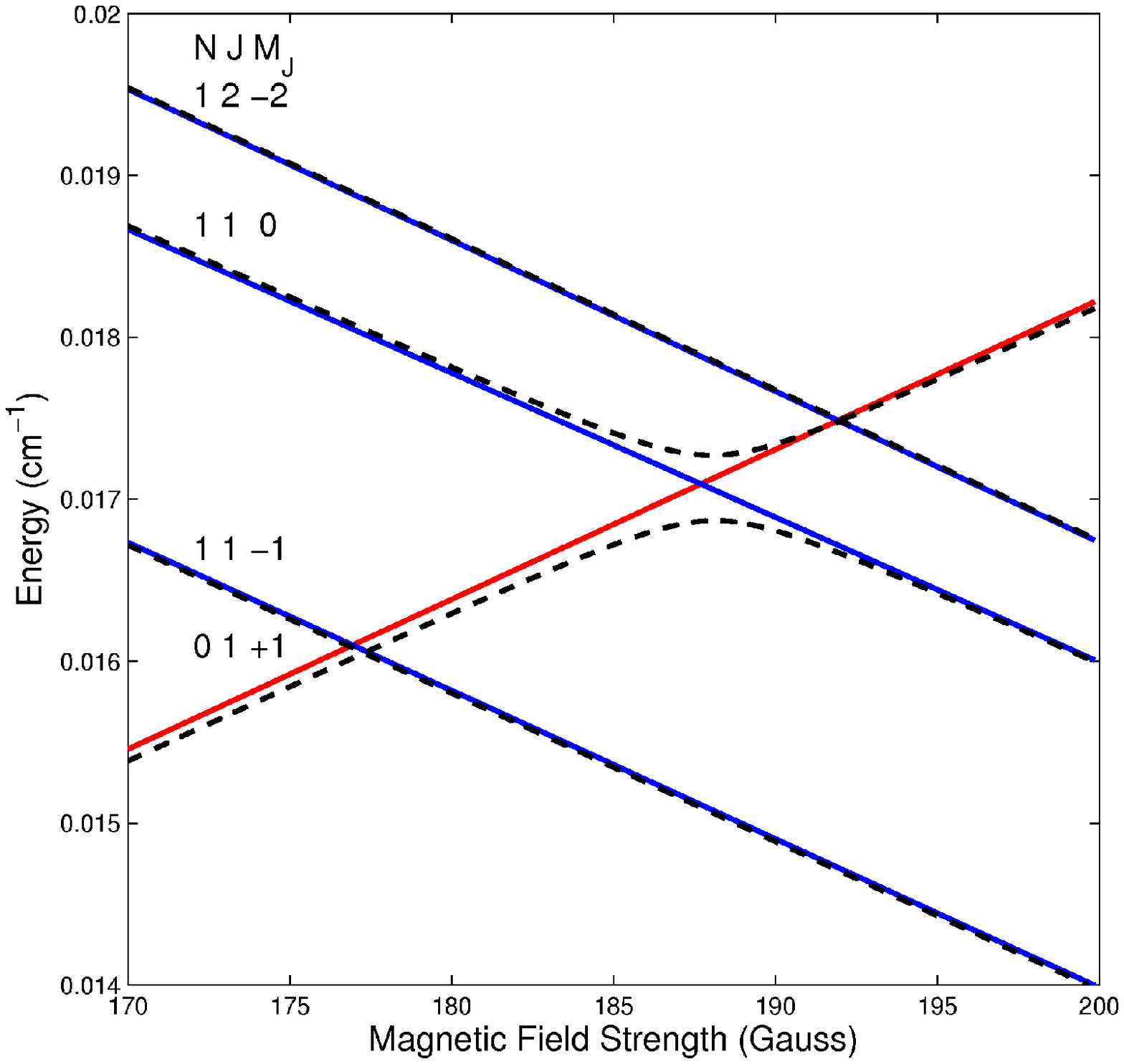}
\caption{ (color online). Energy levels of KRb as a function of
magnetic field $B$. The upper panel shows an overview, with
each zero-field level (labeled by ($N$,$J$) splitting into
$2J+1$ components as a function of magnetic field. The lower
panel shows how two levels of different parity cross ($M_J =$ 0
and +1) in the absence of an electric field (solid lines) but
avoid one another in the presence of a 5 kV/cm electric field
(dashed lines).} \label{figlevs}
\end{figure}

Fig.\ \ref{figlevs} shows the lowest rotational levels of KRb
(a$^3\Sigma^+$) as function of magnetic field in the presence
and absence of a moderate electric field (5 kV/cm). At zero
field, the $N=0$ level has a single sublevel with total angular
momentum $J=1$, while the $N=1$ level is split into 3 sublevels
with $J=0$, 1 and 2. When a magnetic field is applied, each
sublevel is split into $2J+1$ components labeled by $M_J$. At
zero electric field, the $M_J=+1$ level originating from $N=0$,
$J=1$ and the $M_J=0$ level originating from $N=1$, $J=1$ have
different parity and cross near $B_Z=187$~G.

When a non-zero electric field is introduced, parity is no
longer conserved. However, if the electric and magnetic fields
are parallel, $M_J$ is conserved and $M_J = $ 0 and +1 states
cross. We have therefore chosen the electric field to be
perpendicular to the magnetic field to induce an avoided
crossing between the $M_J=0$ and $+1$ states as shown in Fig.\
\ref{figlevs}.

\begin{figure*}[tb]
\subfigure{\includegraphics[width=68mm]{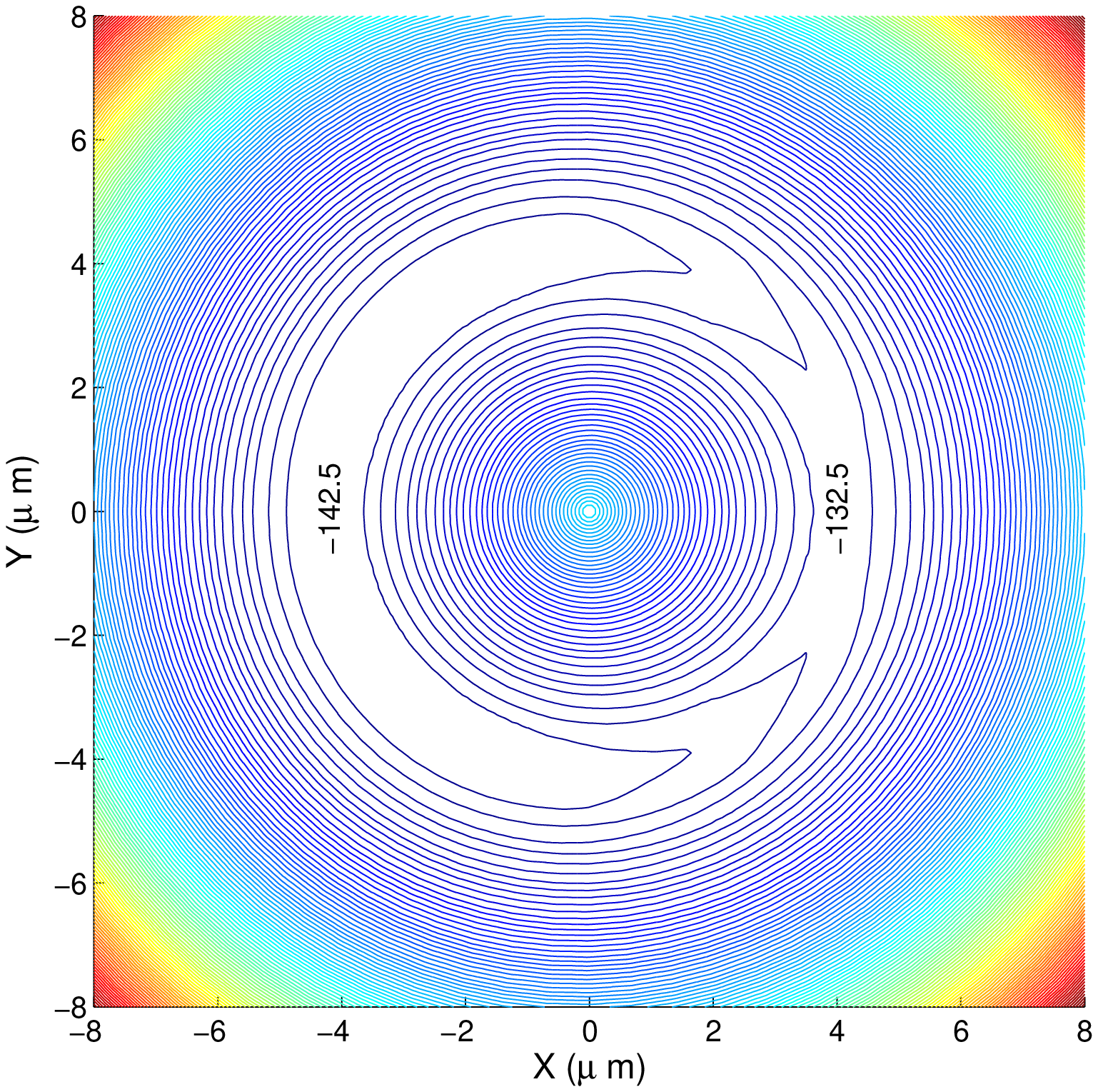}}
\subfigure{\includegraphics[width=68mm]{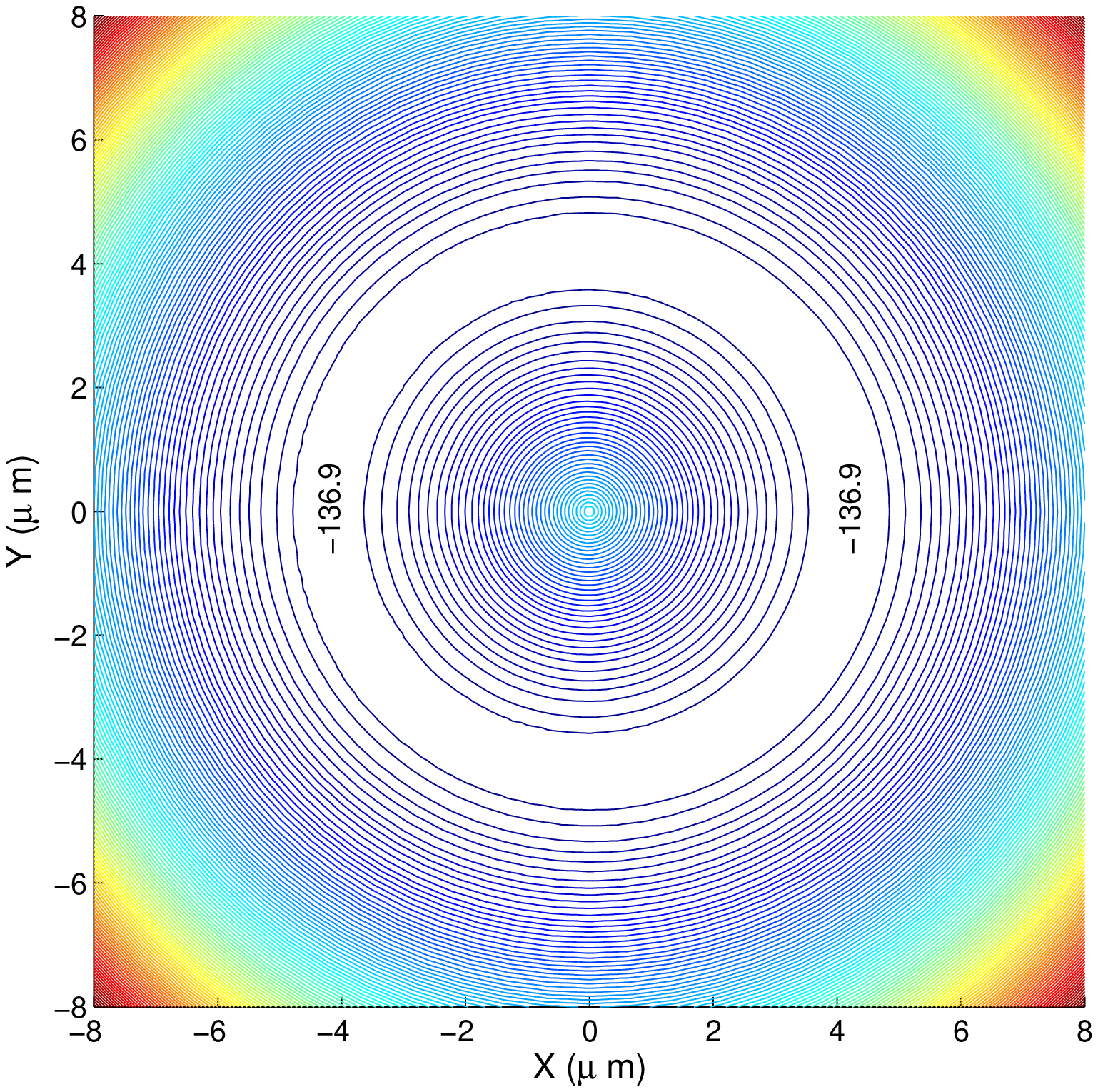}}
\subfigure{\includegraphics[width=68mm]{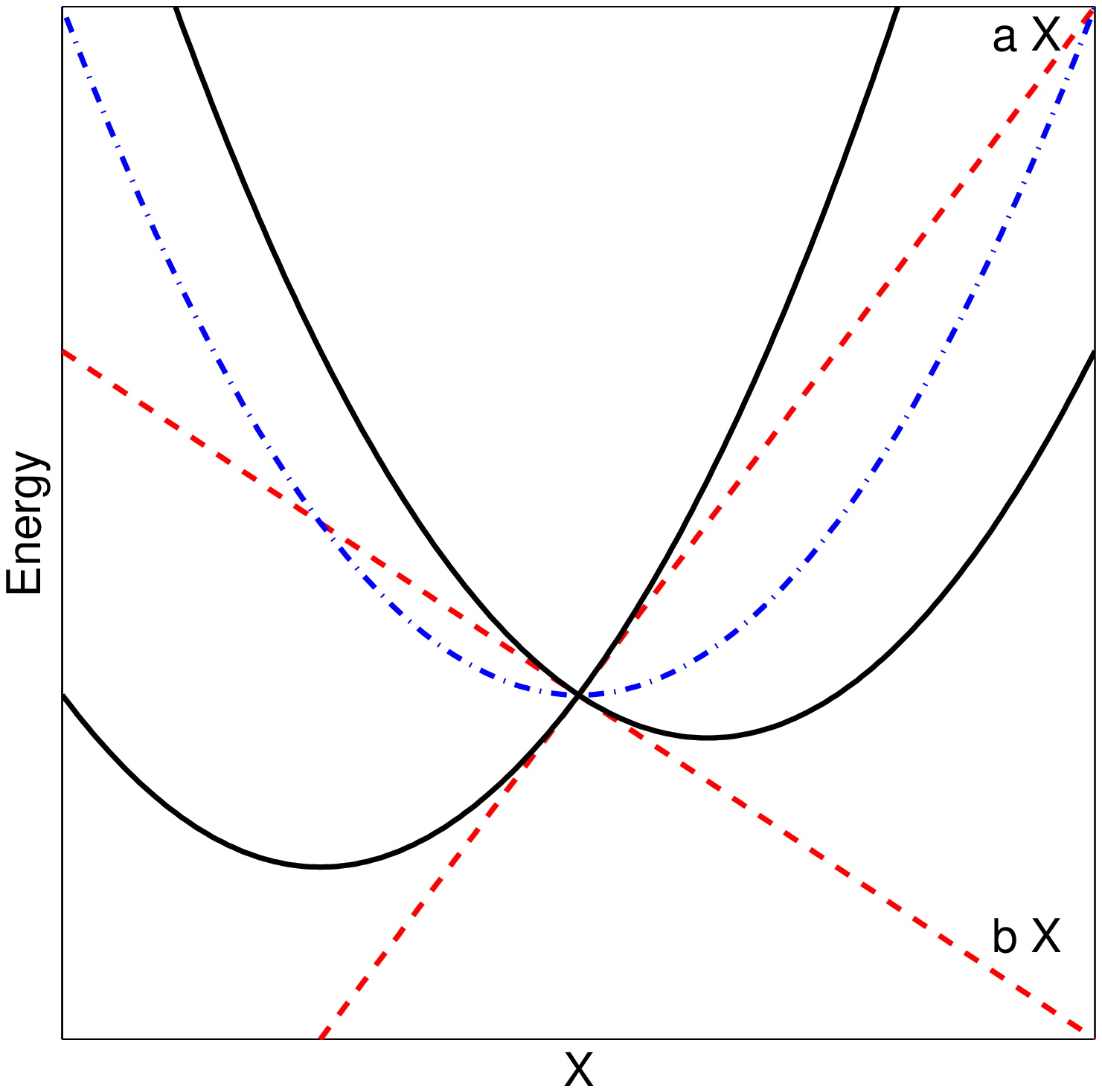}}
\subfigure{\includegraphics[width=68mm]{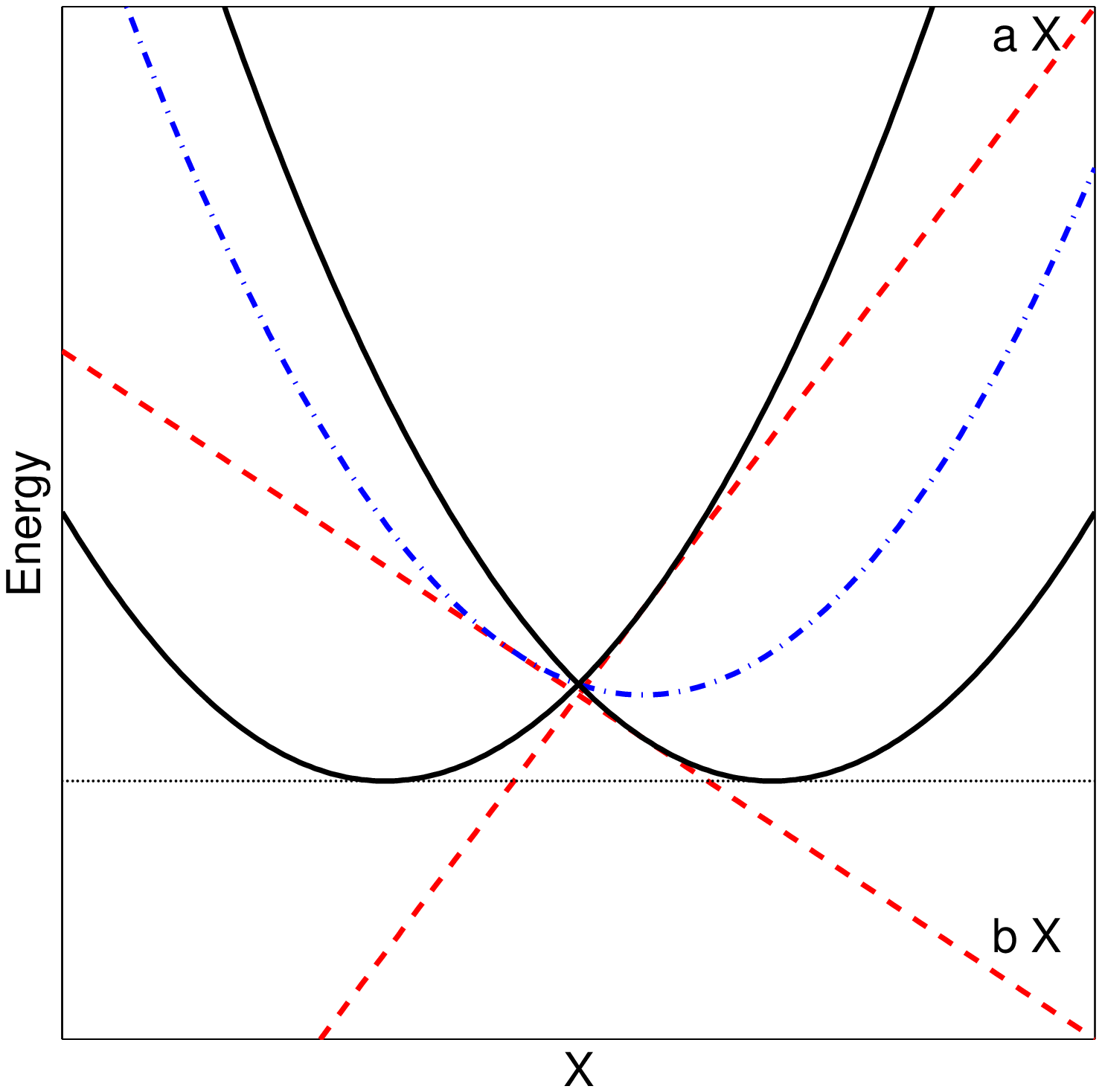}}
\caption{(color online). Toroidal potentials formed around
conical intersections. Left-hand panels: potential formed when
the optical trapping potential (7 nK at 30 $\mu$m) is centered
at the point of intersection ($\rho$=0), with field gradients 5
G/cm $\mathbf{\hat{X}}$ and 6.8 kV/cm$^2$ $\mathbf{\hat{Y}}$.
Right-hand panels: potential formed when the trapping potential
is offset along $\mathbf{\hat{X}}$ by $x_0$ = 0.0715 $\mu$m,
with field gradients 5 G/cm $\mathbf{\hat{X}}$ and 6.723
kV/cm$^2$ $\mathbf{\hat{Y}}$. The electric field gradient is
chosen to minimize the anisotropy in each case. The well depths
are given in nK relative to the point of intersection. The
lower panels show schematic cuts through the potential at
$Y=0$: KRb eigenstates (red, dashed), optical trapping
potential (blue, dot-dashed) and the resultant toroidal
potential (black, solid).} \label{fignooff}
\end{figure*}

We may envisage an experiment in which a BEC is subjected to a
magnetic field $B_Z$, orientated along the space-fixed $Z$
axis, which varies along the $X$ axis with field gradient
$dB_Z/dX$. An inhomogeneous electric field $\bm{E}$ is oriented
along the $X$ axis with a magnitude which varies with $Y$ as
$d\bm{E}/dY$, vanishing on a plane at $Y=0$. This creates a
seam of conical intersections along the line $0,0,Z$ where
$X=0$ is the position at which the magnetic field brings the
two states into degeneracy. Adding an external cylindrically
symmetric optical trapping potential in the $XY$ plane of the
electromagnetic field gradients, $V_{\rm
opt}\left(\rho,\phi,Z\right) = \frac{1}{2}
M\left(\omega_\rho^2\rho^2+\omega_Z^2Z^2\right)$, where $\rho^2
= X^2+Y^2$, creates a toroidally shaped potential around the
conical intersection, with a radial minimum at $\rho_0$. The
left-hand side of Figure \ref{fignooff} shows the resulting
potential for a magnetic field gradient of 5 G/cm, an electric
field gradient of 6.8 kV/cm$^2$ and an optical trapping
potential with a height of $7\ \mu$K at $\rho=30\ \mu$m
\cite{Takekoshi:1998} centered at the conical intersection.
This potential has significant anisotropy (about 10 nK), which
is manifested as an asymmetry along a cut with $Y=0$ as shown
schematically in the lower panel.

The anisotropy of the toroidal trapping potential $V_{\rm
trap}(\phi)$ can be controlled by offsetting the optical
trapping potential from the point of intersection. Assuming
that the KRb eigenstates are linear as a function of magnetic
field over the range of the intersection, with respective
gradients $a$ and $b$, the asymmetry along $Y=0$ will be zero
when the optical trap is centered at $x_0 =
(a+b)/(2M\omega_\rho)$. The trapping potential still has
slightly different depths along the $X$ and $Y$ axes, but this
can be minimized by adjusting the electric field gradient. The
right-hand side of Figure \ref{fignooff} shows the resulting
optimized trapping potential, with $x_0$ = 0.0715 $\mu$m and
$d\bm{E}/dY=6.723$ kV/cm$^2$. This potential has an angular
anisotropy on the order of 0.01 nK.

For a molecule of mass $M$ moving in a toroidal potential such
as those in Fig.\ \ref{fignooff}, the single-particle
Schr\"{o}dinger equation is approximately separable, with
solutions $\Psi(\rho,\phi,Z)=\psi(\rho)\Phi(\phi)\varphi(Z)$.
In the absence of anisotropy, the geometric-phase-induced
anti-periodic boundary condition $\Phi(\phi)=-\Phi(\phi+2\pi)$
gives $\Phi(\phi)=(2\pi)^{-1/2}\exp(i m \phi)$, where $m$ takes
half-integer values $\pm\frac{1}{2}, \pm\frac{3}{2},
\pm\frac{5}{2}$, etc. If the boundary conditions were periodic
over $2\pi$, $m$ would take integer values $0,\pm1,\pm2,$ etc.
For periodic boundary conditions the ground state has zero
angular momentum $m=0$, but for anti-periodic boundary
conditions it has a nonzero angular momentum $m =
\pm\frac{1}{2}$. The single-particle energy spectrum is
\begin{equation}
E = \left(\nu_Z+\frac{1}{2}\right)\hbar\omega_Z+
\left(\nu_\rho+\frac{1}{2}\right)\hbar\omega_\rho
+b_{\rm rot}\left(m^2-\frac{1}{4}\right),
\end{equation}
where the third term represents the rotational energy of the
particle with a rotational constant $b_{\rm rot} =
\hbar^2/2M\rho_0^2$.

In order for geometric phase effects to be observed, the
angular anisotropy of the toroidal trapping potential $V_{\rm
trap}(\phi)$ must be small enough to allow the wavefunction to
fully encircle the intersection. For the potentials shown in
Fig.\ \ref{fignooff}, $\frac{1}{2}\hbar\omega_\rho$ $\approx$ 4
nK and $b_{\rm rot}$ $\approx$ 0.5 nK. For the potential on the
left-hand side, the anisotropy is large compared to $b_{\rm
rot}$, so that the single-particle wavefunction will be
localized on one side of the trap. However, for the potential
on the right-hand side, the anisotropy is small compared to
$b_{\rm rot}$ and the single-particle wavefunction will fully
encircle the conical intersection and exhibit half-integer
quantization.

A BEC of a dilute gas may be modeled by the Gross-Pitaevskii
equation (GPE).  The time-independent GPE is
\begin{equation}
\left[ \hat{H}_0
 + u(\phi)|\Psi(\rho,\phi,Z)|^2\right]\Psi(\rho,\phi,Z) = \mu\Psi(\rho,\phi,Z),
\end{equation}
where $\mu$ is the chemical potential and the mean-field
wavefunction $\Psi$ is normalized to unity, $\int \Psi^*\Psi
d\tau = 1$. The effective interaction strength $u(\phi)$ is
$4\pi\hbar^2N a(\phi) / M$, where $N$ is the number of
particles in the condensate. The internal molecular
wavefunction at angle $\phi$ may be written in terms of the
individual molecular states $\psi_1$ and $\psi_2$ as
\begin{equation}
\psi(\phi)=\psi_1\cos(\phi/2)+\psi_2\sin(\phi/2),
\label{eq:psiphi}
\end{equation}
so the effective angle-dependent scattering length $a(\phi)$ is
\begin{eqnarray}
a(\phi) =&\textstyle{\frac{1}{8}}&(3a_{11}+3a_{22}+2a_{12})
+ \textstyle{\frac{1}{2}}(a_{11}-a_{22}) \cos\phi \nonumber\\
+ &\textstyle{\frac{1}{8}}&(a_{11}+a_{22}-2a_{12})\cos2\phi,
\label{eq:aphi}
\end{eqnarray}
where $a_{ij}$ is the scattering length for interaction between
molecules in states $i$ and $j$. This is isotropic in the case
$a_{11}=a_{22}=a_{12}$.

Averaging over the radial and vertical wavefunctions gives an
effective 1D GPE in $\phi$,
\begin{equation}
\left[ -b_{\rm rot}\frac{\partial^2}{\partial\phi^2} +
\tilde{u}(\phi)|\Phi(\phi)|^2\right]\Phi(\phi) = \tilde{\mu}\Phi(\phi),
\end{equation}
where
\begin{equation}
\tilde{u}(\phi) = \frac{4\pi\hbar^2N a(\phi)}{M}
\iint|\psi(\rho)|^4|\varphi(Z)|^4\rho\,d\rho\,dZ
\label{eq:uphi}
\end{equation}
and $\mu = \tilde\mu + \left(\nu_Z+\frac{1}{2}\right)
\hbar\omega_Z+\left(\nu_\rho+\frac{1}{2}\right)\hbar\omega_\rho
-b_{\rm rot}/4$. For a flat ring, this has analytical solutions
$\Phi(\phi)=(2\pi)^{-1/2}\exp(im\phi)$, with $\tilde{\mu} =
b_{\rm rot}m^2 + \tilde{u}/(2\pi)$. Applying antiperiodic
boundary conditions, we obtain the same solutions as for the
single-particle case, namely states of half-integer quantized
angular momentum.

In the presence of a small anisotropy, there are two classes of
solution satisfying antiperiodic boundary conditions: flowing
solutions, $\Phi_m^\pm(\phi) \approx (2\pi)^{-1/2}\exp(\pm
im\phi)$, and static solutions, such as $\Phi_m^0(\phi) \approx
\pi^{-1/2}\cos(m\phi)$ for small interactions, both with
half-integer $m$. For a trap with a residual anisotropy $V_{\rm
trap}(\phi) = -V_1\cos\phi-V_2\cos2\phi$ and an angle-dependent
interaction strength $\tilde u(\phi) = u_0 + u_1\cos\phi +
u_2\cos2\phi$, we obtain approximate chemical potentials
corresponding to these two Ansatzes of $\tilde\mu_{1/2}^\pm
\approx b_{\rm rot}/4 + u_0/(2 \pi)$ and $\tilde\mu_{1/2}^0
\approx b_{\rm rot}/4 -V_1/2 + 3u_0/(4 \pi) - u_1/(2 \pi) -
u_2/(8 \pi)$. In this approximation, a flowing state with
$m=\pm1/2$ is the ground state if $2u_1 + u_2/2 + 2 \pi V_1 <
u_0$. For any given $a(\phi)$ and condensate number, we can
apply an offset of the optical potential sufficient to
compensate for the anisotropy of the interaction term and
stabilize the flowing state.

The permitted velocities of the persistent flow can assume only
half-integer values compared to the quantized units of
circulation possible in a more conventional single-species
atomic BEC in a comparable toroidal geometry \cite{ryu2007}.
The persistent flow should be observable by releasing the
trapped particles and employing a time-of-flight technique
\cite{ryu2007}.

A BEC such as described here is stable only if $a(\phi)$
remains positive all around the ring. From Eq.\
(\ref{eq:aphi}), this requires that $a_{11}$ and $a_{22}$ are
both positive and that $2a_{12} > -(a_{11} + a_{22})$. In
addition, a condensate of polar molecules can undergo dipolar
collapse \cite{Koch:2008} if the dipole length $a_{\rm d}$
exceeds the scattering length for the short-range interactions,
where $a_{\rm d}=|d^2| M / (12 \pi \epsilon_0 \hbar^2)$ and $d$
is the effective dipole moment of the molecule in the field.
Since the molecular wavefunction is given by Eq.\
\ref{eq:psiphi} and there is a direct dipole moment matrix
element $\langle1|\mu|2\rangle$ between the near-degenerate
states $\psi_1$ and $\psi_2$,
$d=\langle1|\mu|2\rangle\sin\phi$. For the two states of KRb
considered here, $\langle1|\mu|2\rangle \sim 10^{-32}$~C\,m,
which gives $a_{\rm d}^{\rm max} \sim 5\times 10^{-12}$~m. This
is substantially smaller than typical scattering lengths so
dipolar collapse is unlikely.

The present Letter has described a novel form of conical
intersection that can occur as function of 3-dimensional
laboratory coordinates, instead of internal molecular
coordinates. Using such conical intersections, it may be
possible to create novel superfluid states with stable
persistent flow characterized by half-integer, rather than
integer quantized angular momentum. Although we have considered
the effect for polar molecules, a similar effect might be
produced for well-separated levels, perhaps even in atoms,
using a laser field to bring the levels into near-degeneracy
and an inhomogeneous magnetic field to provide a crossing.
Conical intersections would appear at points of zero laser
amplitude, for example at the nodes in an optical lattice.

The effect proposed here can in principle be observed in any
system where two states of any different symmetry (not just
parity) can be tuned into degeneracy with one external
influence, and then split apart again with another influence
that breaks the symmetry.

The authors are grateful to EPSRC for a PhD studentship for
AOGW and for funding the collaborative project QuDipMol under
the ESF EUROCORES programme EuroQUAM.

\bibliography{all,sg}

\end{document}